# Pathway: a protocol for algorithmic pricing of a DAO governance token

GC DAO
info@gton.capital

| Aleksei Pupyshev | Ilya Sapranidi | Shamil Khalilov |
| alexp@gton.capital | ilyas@susy.one | shama@susy.one |

## Abstract

In this paper, we will consider a governance token pricing algorithm that conducts liquidity operations on AMM CPMM DEXs with liquidity that belongs to a decentralized autonomous organization (DAO), also called protocol-owned liquidity (POL). The primary aim of the protocol is maintaining a price peg by determining algorithmically when and how to carry out interventions that consist of two steps: extracting liquidity from an AMM liquidity pool and conducting "token swap" operations. We will cover setting up an optimal peg function as a weighted sum of certain normalized factors, which are to be determined collectively by the DAO. In particular, we will review *various arithmetic invariants of liquidity intervention, which brings the price to a peg while leaving total liquidity intact, and show how such interventions can be substituted in practice by so-called PMM (proactive market maker) protocols.*

In addition, we will demonstrate how a systematic application of Pathway protocol by DAOs can create a class of so-called algorithmic governance tokens (AGT). In the end, we will consider ways of technical implementation of Pathway using Solidity, as well as potential practical problems that may occur when implementing such a protocol on EVM (Ethereum virtual machine) blockchains and AMM DEXs. Implementations of PMM protocols that Pathway can be built upon are included in the reference section.

## Problem

In the traditional, passive approach to DAO-operated liquidity, the market value and the intrinsic value of a DAO token are correlated weakly. The market value stability is disrupted by external noise, intricacies of traders' behavior, as well as incomplete or asymmetric information. Passive market-making strategies employed by DAOs are prone to the influence of feedback loops, where small changes in indicators can result in market panic. Algorithmic DAO tokens, on the other hand, can use the approach presented in this paper in order to smooth and stabilize the price of the token by linking it to the intrinsic value peg. We believe that the implementation of such mechanics can increase market efficiency and prosperity of DAO systems and their token holders through establishing positive feedback loops that correlate with the fundamental growth factors of a token.

We assume that for a DAO that owns protocol liquidity, placing a token on a classic AMM is problematic, since it is almost impossible to establish a direct connection between a project's performance and its pricing. There are also issues that involve large market makers and token holders with the ability to influence the market, as well as insider trading, non-transparent MM processes, and others in which there is a strong imbalance in the distribution of the governance power held over the token. Moreover, passive AMM strategies of DAOs, where liquidity is provided into pools but never managed nor rebalanced, can lead to uncontrolled bursts of token volatility, which we interpret as violation of game-theoretic Nash equilibria.

The Pathway protocol can become a tool that DAOs can use to formalize a set of fundamental factors that influence the pricing of a token and establish a market making system that amplifies and sustains this correlation.

## Definitions

**DEX (decentralized exchange)** is a peer-to-peer marketplace where transactions occur directly between traders or between traders and smart contracts without interference of a counterparty or a centralized entity.

**An MM (market maker)** is an open market participant which is providing assets for both the demand and supply sides according to certain trading strategies, taking into account asset pricing models on exchanges.

**An AMM (automated market maker)** is a type of DEX that relies on a mathematical formula to price assets. The formula is implemented as a smart contract and can vary with each protocol. A classic approach pioneered by Uniswap and used by multiple AMM DEXs [1] is CPMM (constant product market maker) which defines the price relation between two tokens based on their amounts in the pool.

**CPMM Formula** is as follows:

$$Amount(G) \cdot Amount(U) = C,$$

where $Amount(G)$ is the supply of G token in the AMM pool, $Amount(U)$ is the supply of token U in the pool, C is constant.

**Token swap operations** conducted with the pool pair change the ratio between $G$ and $U$, and in doing so, the price changes, while $C$ remains the same. Thus, the price of one $G$ token in $U$ tokens is always defined as:

$$Price(G) = \frac{Amount(U)}{Amount(G)}$$

$$Price(G) = \frac{C}{Amount(G)^2}$$

**LP (liquidity provision) tokens** are tokenized shares of funds in AMM pools represented as $G$ and $U$ tokens pair locked on a smart contract. Extracting liquidity means "burning" LP tokens with the release of $G$ and $U$ in the proportion defined via the CPMM formula.

Adding or removing liquidity to the pool will change $Amount(G)$ and $Amount(U)$, which should also change C according to the CPMM formula.

AMM liquidity in the pool measured in tokens can be defined as:

$$Liquidity(U) = Amount(G) \cdot Price(G) + Amount(U)$$

**DAO (Decentralized Autonomous Organization)** is an organizational structure of members that share a common aim without the need of a centralized controlling entity and operating on the smart contracts. The use of blockchain allows DAOs to define (governing) rules and manage a project treasury in both a transparent and secure way, by creating, accepting and rejecting DAO proposals according to voting results.

**POL (protocol owned liquidity)** describes an approach to liquidity management where LP-tokens are owned and managed collectively by a DAO. It is used to bypass a common problem in DeFi which is a constant need for attracting and retaining liquidity for the token. The POL concept was pioneered by Olympus DAO. [2]

**Intrinsic value** is a measure of what an asset is worth. This measure is arrived at by means of an objective calculation or complex financial model, rather than using the currently trading market price of that asset. [3] In this document, intrinsic value will be called "peg price" or simply peg.

**Market price** is the current price at which an asset or service can be bought or sold. The market price of an asset or service is determined by the forces of supply and demand. The price at which quantity supplied equals quantity demanded is the market price.

**Proactive Market Maker (PMM)** is a novel blockchain market making model. With PMM, parameters such as asset ratio and curve slope that determine the mathematical relationship between assets can be set flexibly. To guide prices and conduct market price discovery on-chain, an oracle can be introduced into a PMM system. [8]

**Decentralized Oracle Networks (DONs)** are infrastructure networks for "pushing" price, peg or any other data feeds into smart contracts. According to the Chainlink white paper [4], "A DON is a network of oracles maintained by a committee of Chainlink nodes. It supports any of an unlimited range of oracle functions chosen for deployment by the committee. A DON thus acts as a powerful abstraction layer, offering interfaces for smart contracts to extensive off-chain resources and highly efficient yet decentralized off-chain computing resources within the DON itself."

# Peg Models

There are multiple approaches to estimating what intrinsic value should be attributed to a governance token.

## Portfolio peg

A DAO can manage a portfolio of assets that consists of various tokens and regularly conduct rebalancing. In this case the governance token is essentially an index of this portfolio. A similar approach can work with NFT portfolios and portfolios with mixed asset combinations.

A trivial approach that such a "portfolio-oriented DAO" can use to establish a peg to a certain value of the governance token is as follows:

$$PWPeg(t) = \frac{\sum_i c_i \cdot FV(t, A_i)}{GovTokenSupply(t)},$$

where $FV$ is future value [5], which is defined as the price of the governance token at a future date based on an assumed rate of growth, $c_i$ is the amount of asset $A_i$ owned by the DAO.

## Linear model

If the business model of the DAO is not limited to managing and balancing a portfolio of assets, but is oriented towards achieving some goals set for development of infrastructure and products and to increase success metrics of those products and the ecosystem as a whole, a more flexible model based on a set of factors can be applied.

To do so, the DAO has to convey to determine collectively:

1. a list of numerical factors $F_{t,i}$ that reflect the success of the DAO, for instance: liquidity, active users, integrations, transactions, revenue, assets under management.

2. impact coefficients (weights) $w_i$ of these factors, such that:

$$\sum_i w_i = 1$$

3. A normalizing function for the factors which vary over time $t$, that normalizes each factor to the interval [0,1] , for example:

$$\hat{F}_{t,i} = \frac{F_{t,i} - Min(F_{t,i})}{Max(F_{t,i})}$$

After normalizing factors, multiplication with weights will produce prices between 0 and 1. Therefore, a scaling function $S$ should be introduced that maps the sum of weighted factors onto an interval $[0, MaxPrice]$, where $MaxPrice$ is the ceiling value of the governance token price determined by the DAO.

Also, in order get rid of determinism, which can give rise to arbitrage or front-running attacks, a random noise should be introduced into the system.

Thus, a peg for the pricing of governance tokens can be represented as a weighted sum of normalized factors with added noise:

$$PWPeg(t) = bias + S(\sum_i w_i \hat{F}_{t,i}) + noise(t),$$

where $t$ is time, $bias$ is a floor price value parameter (determined by the DAO), S is a scaling function and $noise(t)$ is random noise.

To some practically acceptable degree of accuracy, all possible models can be described using the linear model.

## Intervention

Intervention is a liquidity operation procedure initiated by a DAO MM, which is aimed at adjusting the market price towards the peg (intrinsic value) price by conducting operations with POL. In this example, U is a stablecoin (a quote token), and G is a governance token of the DAO.

Let us analyze a numerical example:
1. Consider an AMM pool with $20U + 10G$ tokens.

   Its state is:
   $Price(G) = 2$ (nominated in U tokens),
   $Amount(U) = 20$,
   $Amount(G) = 10$

   Total liquidity in the pool in U tokens is:
   $Liq(U) = Amount(U) + Amount(G) \cdot Price(G) = 40$

2. The new peg price is:

   $Peg(G) = 3$

   Therefore, an intervention is required to adjust $Price(G)$ towards $Peg(G)$:

   $Price(G, S_1) = 2$,
   $Price(G, S_2) = Peg(G) = 3$,

   where $S_1$ and $S_2$ are State 1 and State 2 before and after an intervention correspondingly.

3. In the perfect scenario, all assets contained in the AMM pool are owned by the DAO. This can also mean that no extra funds in $U$ token can be made available to change the market price. If there are no extra funds at DAO's disposal (e.g. from the treasury), $Amount(U)$ must remain the same after Up-Intervention. Therefore, an approach is needed that avoids using any extra funds to conduct the swap, extracting the necessary amount from liquidity instead:

   $Liq(U, S_2) = Liq(U, S_1) = Amount(U) + Amount(G, S_2) \cdot Peg(G)$

   According to the CPMM formula, $Amount(G, S_2)$ must be equal to 6.67 $G$ tokens in State 2. In this case, $Liq(U)$ remains unchanged i.e. equal to 40 before and after the Up-Intervention. Let us examine how the MM should approach determining the amount to withdraw from the pool.

4. In pools on AMM DEXs, the total amount of liquidity the user is able to remove is equal to their share represented as an LP token. LP tokens essentially represent a pair of *G* and *U* tokens. Due to the structure of UniV2 AMM CPMM pools, it is impossible to remove an asset from one side of the pool: *G* and *U* tokens can only be withdrawn simultaneously in the proportion fixed by the CPMM formula.

In order to execute an Up-Intervention, the MM must withdraw (burn) X LP tokens from the pool and swap *U* tokens into *G* tokens through the same pool to adjust the price. Thus, X value in *G* tokens is calculated as follows, with some approximation:

$Amount(U) + Amount(G, S_2) \cdot Peg(G) = Liq(U, S_2)$
$20 + 6.67 \cdot 3 \approx 40$
$Amount(G, S_1) - Amount(G, S_2) = X$
$10 - 6.67 = 3.33G$

Removing 3.33 G in LP tokens means that the number of the extracted *G* tokens is equal to $3.33/2 = 1.665G$ tokens and the number of the extracted U is 1.665 nominated in G tokens.

At this point, the DAO MM should have a sufficient amount of U tokens on the balance for conducting the price intervention. To do so, the *X* liquidity should be removed, tokens *U* should be swapped into tokens *G*, with *3.33G* tokens remaining on the MM balance, resulting in a new pool state with 6.67 G and 20 U tokens. $Liquidity(U)$ is still 40 (nominated in U) and $Price(G, S_2)$ is 3.

5. As a result, such Up-Intervention will increase the market price of G token (2→3), occurring due to the removal of 6.67 G tokens from the pool.

**State changes**
Pool before Up-Intervention:
$S_1$
$20U + 10 \cdot Price(G) = 40U$
$Price(G, S1) = 2$
$Balance(G, S1) = 0$
↓
Pool after Up-Intervention:
$S_2$
$20U + 6.67 * Price(G) \approx 40U$
$Price(G, S_2) = 3$
$Balance(G, S_2) = 3.33$

Due to the symmetry of the CPMM formula, Down-Intervention, which might be needed to retain the price in a certain corridor, is identical to Up-Intervention, except for changing tokens *U* and *G* and redefining the prices.

# Formalization

Let us define:
- $P_1$ as $Price(G, S_1)$, $P_2$ as $Price(G, S_2)$ i.e. price of G tokens nominated in U tokens from the example above;

- $S_1 \to S_2$ as Up-Intervention, $P_1 \to P_2$ as adjusting the market price to the peg price as a result of $S_1 \to S_2$ intervention;
- $Amount(G, S_1)$ as $g$, $Amount(G, S_2)$ as $g'$, $Amount(U, S_1)$ as $u$ and $Amount(U, S_2)$ as $u'$;
- *X* as amount of LP tokens which has to be extracted from AMM pool in order to accomplish Up-Intervention. The proportions of tokens *U* and *G* extracted in the form of LP tokens can be defined as $X_u$ and $X_g$;
- The amount of *G* tokens received as a result of swapping $X_u$ into *G* tokens can be defined as $S_g$.

### Thesis 1

To execute a $P_1 \to P_2$ intervention, the DAO should algorithmically remove the excess amount of the governance token (determined as described in the "Intervention" section), while the quantity of the market token in the pool should remain the same. This upward movement is thus called Up-Intervention. Vice versa, when making a Down-Intervention, an excess of the quote token should be removed. Both cases are symmetrical due to the nature of CPMM formula.

### Thesis 2

The DAO intervention should not change the total liquidity value of the pool. That is, the sum of the pair of tokens in the pool should remain the same after the intervention, regardless of its direction:

$$g \cdot P_1 + u = g' \cdot P_2 + u',$$

where $u = u'$, i.e. the number of tokens *u* does not change as a result of the intervention.

Furthermore:

$$g' = \frac{g \cdot P_1}{P_2}$$
$$\Delta g = g' - g = g \cdot (\frac{P_1}{P_2} - 1)$$

This formula demonstrates that to change $P_1 \to P_2$, one needs to withdraw $|\Delta g|$ tokens.

Due to the symmetry of the formulas, similar calculations are also valid for the *U* token.

Therefore, the intervention should take three steps:

I. Determine the amount of liquidity to be extracted from the pool (*X* LP tokens, LPs);
II. Burn X LPs with "unwrap" the pair of tokens: $X \to X_g + X_u$;
III. Swap $X_u$ to $X_g$, which leaves $2X_g$ tokens, which is equal to $|\Delta g|$ tokens in the formula above.

## Thesis 3

In a CPMM system that is isolated and contains $|\Delta g|$ tokens after the intervention, the exact algorithm that led to this result does not impact any value, except P1 changing into P2 and the number of *G* tokens in the pool. That is, the result of the intervention does not depend on how exactly it was carried out, which means that any strategy that leads to $P_1 \rightarrow P_2$ while maintaining an invariant pool value is practically applicable.

The remainder $|\Delta g|$ contains *G* token from two sources: LP ($X_g$) and Swap ($X_u \rightarrow S_g$):

$$|\Delta g| = X_g + S_g$$

By virtue of Thesis 3, $X_g = S_g$, meaning that half of the *G* tokens can be gained from LP "burn", and the remaining half by swapping ($X_u \rightarrow S_g$).

Based on this thesis, we can describe the Up-Intervention algorithm as such:
I. Determine $P_2$ and calculate *g*, *u*, and $P_1$;
II. Compute $\Delta g = g \cdot (\frac{P_1}{P_2} - 1)$ and the *X* amount of LP to withdraw which consists of $\frac{|\Delta g|}{2}$ tokens;
III. Remove *X* LP and unwrap $X_g$ and $X_u$ tokens;
IV. Swap all the removed $X_u$ to *G* token to receive $\frac{|\Delta g|}{2}$ G tokens.

This is a formalization of the procedure that conducts an Up-Intervention. A Down-Intervention works similarly due to the symmetry of the CPMM AMM formula.

## Implementation issues

One of the difficulties in the practical implementation of Pathway is related to liquidity ownership. For the protocol to perform successful interventions, the DAO should own most of the liquidity for its token. The proportion of POL determines the level of control that the DAO has when attempting to achieve a peg. The exact way of structuring the DAO to ensure liquidity accumulation in the form of POL can be implemented in various ways which are not a subject of this paper. One of such ways is bonding, that is, a sale of staked governance token allocations which become an integral part of the governance token liquidity.

Another important requirement is the security and reliability of oracles that provide factors, peg and market price data.

Perhaps the most important issue we see that can influence Pathway's deployment is front-running transactions that can undermine DAO interventions. In this case, we envision two practical strategies for protection.

With the first strategy, a threshold for the price difference from the oracle and the AMM pool is established at the time of execution. This is analogous to slippage tolerance in AMM DEXs. However, we consider this kind of protection strategy comparatively weak.

In order for the protection to be sufficiently strong, a VRF (verifiable random function) randomizer should be employed at the time of intervention execution, which makes it known whether the transaction will be executed or rejected only at the time of execution. [6] With this strategy implemented, front-running will be more risky for the frontrunner, as it does not guarantee profit but forces the frontrunner to enter a token position.

If the number of market makers who arbitrage against Pathway increases, the number of required liquidity interventions will decrease accordingly, since, in fact, those market makers will likely outpace the interventions while competing with each other.

There is an important limitation of Pathway interventions that arises when the peg price is constant or stable. We call this an "intervention sandwich" attack.

Imagine Alice sold *A* tokens in the pool. Bob then buys back $\frac{A}{2}$ tokens and waits for a Pathway liquidity intervention to occur. As soon as it happens, Bob can immediately sell his $\frac{A}{2}$ tokens for a profit. This attack is possible because the price will deterministically return to a stable peg at some point, therefore Bob can earn a profit risk-free, which leads us to assuming that Pathway interventions on univ2-type pools are flawed when using a stable peg.

To prevent such "intervention sandwich" attacks, we believe Pathway should either be implemented on a protocol-owned AMM DEX with an intervention being executed for each swap, or a so-called Proactive Market Maker (PMM) DEX can be used to maintain the peg. However, both of these solutions have a number of issues related to cross-market arbitrage, where one pool is based on PMM or an instant Pathway AMM, and another one is a classical CPMM pool, an AMM DEX pool or even an orderbook-based DEX/CEX. To solve the arbitrage issue, a liquidity concentration incentives model and competitive advantages for arbitrage accounts which belong to the DAO (whitelisted) can be introduced.

## Proactive Market Maker

A universal solution for DAO token pricing which can work both for the stable and volatile price pegs is to implement Pathway on a special type of DEXs based on the PMM (proactive market maker) approach such as DODO EX. An implementation of the DODO PMM can be found in [8].

In DODO EX, the PMM price curve follows this pricing formula:

$$P_{margin} = i\,R,$$

where R is defined as the piecewise function below:

$$if : B < B_0, R = 1 - k_1 + (\frac{B_0}{B})^2 k_1$$
$$if : Q < Q_0, R = 1/(1 - k_2 + \frac{Q_0}{Q})^2 k_2)$$
$$else : R = 1,$$

where $B_0$ is the total number of base tokens, $Q_0$ is the total number of quote tokens, $B$ is the base token balance, $Q$ is the quote token balance, $i$ is the market price provided by an oracle, and $k_1$, $k_2$ are slope parameters in the range [0,1].

In the "Formalization" chapter for CPMM, we showed that after each intervention, the DAO is left with an asymmetric balance of tokens. Classical AMMs cannot operate with asymmetric liquidity, whereas the PMM approach, on the other hand, opens an opportunity to put such extracted liquidity to use by keeping it in the pools and adjusting the PMM bonding curves.

Other benefits of the PMM approach explained in [9] is that although the AMM price is a subject of arbitrage activity, PMM sets the price based on oracle data. In case of a DAO token coupled with a volatile asset which is declining in price, an AMM would bring its price down through arbitrage, but a PMM would not force the selling of the token. This approach can benefit the DAO and open opportunities to support and incentivize pools with volatile assets without a direct negative influence on the DAO governance token.

Some authors [10] [11] suggested similar approaches to PMM and call them dynamic AMM that use so-called dynamic curves and rely on oracles. The proposed approach would utilize input from a market price oracle to adjust the relationship between the assets so that the pool price becomes identical to the market price. As stated in [10], this approach would eliminate arbitrage opportunities and maintain liquidity in the pool for all assets and the total value of the pool over a wide range of market prices.

To summarize, several different potential implementations of Pathway exist: liquidity interventions to maintain a volatile price peg on a Univ2 AMM, or the PMM approach for any types of peg, including stable. Thus, the PMM approach opens an opportunity for DAO-owned stablecoins or mirrored/synthetic assets pegged to any price.

## Summary

The Pathway algorithm presented in this paper allows for a complete formalization and automation of the pricing process for governance tokens of DAOs, which creates a new decentralized financial primitive: algorithmic governance tokens (AGT).